\documentclass[12pt]{article}
\pdfoutput =1
\usepackage{float} 
\usepackage{amsmath}
\usepackage{slashed}
\usepackage{amsfonts}   
\usepackage{amssymb}

\begin{document}
\date{}
\title{{\bf{\Large Multispin magnons on deformed $ AdS_{3}\times S^{3} $}}}
\author{
 {\bf {\normalsize Dibakar Roychowdhury}$
$\thanks{E-mail:  dibakarphys@gmail.com, dibarak@post.bgu.ac.il}}\\
 {\normalsize  Department of Physics, Ben-Gurion University of The Negev,}\\
  {\normalsize P.O. Box 653, Beer-Sheva 84105, Israel}
\\[0.3cm]
}

\maketitle
\begin{abstract}
In this paper, using the holographic prescription, we study multispin bound states and their dispersion relations over the $ \kappa $-deformed $ AdS_{3}\times S^{3} $ background. In the first part of our analysis, considering the conformal gauge conditions (associated with the Polyakov action) we explore the dispersion relation associated with spin two bound states at strong coupling. We solve corresponding world-sheet fluctuations and compute all the conserved quantities associated with the stringy dynamics over the deformed background. In the second part of our analysis, we perform similar analysis for spin three configurations. In both cases, we observe the emergence of non trivial background deformation that vanishes in the limit, $ \kappa \rightarrow 0$.
\end{abstract}

\section{Overview and Motivation}
According to the celebrated $ AdS_{5}/CFT_{4}  $ correspondence \cite{Maldacena:1997re}, the type IIB string theory formulated in $ AdS_{5}\times S^{5} $ background is dual to strongly coupled $ \mathcal{N}=4 $ SYM in four dimensions. Given this astonishing prescription, one could in fact think of several remarkable implications and/or consequences that naturally emerges out of this duality conjecture. One immediate consequence of this duality conjecture turns out to be the obvious equivalence between the spectrum of stringy excitation in $ AdS_{5}\times S^{5} $ to that with the spectrum of operator dimensions in $ \mathcal{N}=4 $ SYM theory. In order to test this duality conjecture, one therefore needs to check the full quantum spectrum on both sides of the duality which is undoubtedly a difficult job in itself. However, it turns out that this situations seems to get quite manageable under certain special circumstances namely, in the limit where the number of colors becomes large, $ N \gg 1 $. This is the so called planar limit of the duality where one could in principle carry out semi-classical computations on the string theory side in order to compare it with the spectrum of anomalous dimensions corresponding to single trace (gauge invariant) operators on the dual gauge theory side. In other words, the theory becomes \textit{integrable} on both sides of the duality \cite{Serban:2010sr}. 

A remarkable breakthrough along this direction came through the proposal due to Minahan and Zarembo \cite{Minahan:2002ve} who sort of unveiled an astonishing connection between spin chains and that of the stringy dynamics in $ AdS_{5}\times S^{5} $ by identifying the Hamiltonian operator corresponding to the spin chain systems to that with the dilatation operator in $ \mathcal{N}=4 $ SYM. Building on these ideas, several proposal came up and various attempts were made \cite{Kruczenski:2003gt}-\cite{Stefanski:2004cw} in order to understand this connection to some deeper extent. 

One remarkable achievement along this particular direction came through the discovery of the underlying connection between the physics of the spin wave (magnon like) excitation associated with long spin chains to that with certain specific (rotating and pulsating) stringy configurations in $ AdS_{5}\times S^{5} $ \cite{Hofman:2006xt}-\cite{Ahn:2011zg}. In the following we elaborate on this issue in a bit detail. We consider the limit where one of the conserved charges ($ J $) of the dual $ SO(6) $ symmetry becomes infinitely large. This is the so called limit where one considers infinitely long chain of single trace operators on one side of the duality and infinitely long strings on the other side that eventually simplifies the computation enormously. On the dual gauge theory side one considers operators with large scaling dimensions namely, $ \Delta \geq J $ such that the difference, $ \Delta -J $ always appears to be finite while keeping the t'Hooft coupling ($ \lambda $) of the theory fixed. On the dual stringy picture one recovers identical picture where both the energy of excitation ($ E $) and the angular momentum ($ J $) of the string becomes infinitely large while keeping the difference finite. 

In order to understand the physics of spin waves associated with spin chain systems more rigorously, one could imagine long single trace operators on the dual gauge theory side which has $ J $ number of operator ($ Z $) insertion in it. This $ Z $ essentially stands for the ground state of the spin chain configuration. In order to add excitation to this spin chain configuration one could imagine adding an operator ($ Y $) from outside and consider all possible operator insertion namely,
\begin{eqnarray}
\mathcal{O}\sim \sum_{k}e^{ikp}(ZZ....ZYZ...ZZZ)\label{O}
\end{eqnarray} 
which eventually corresponds to the propagation of the \textit{magnon} excitation over this infinitely long spin chain.

Using SUSY, it was Beisert \cite{Beisert:2005tm} who first computed the spectrum associated with the above spin chain system and arrived at the following dispersion relation corresponding to these magnon excitation associated with the spin chain configuration,
\begin{eqnarray}
\Delta -J=\sqrt{1+\frac{\lambda}{\pi^{2}}\sin^{2} \frac{p}{2}}\label{M}
\end{eqnarray} 
which in the limit of the large t'Hooft coupling ($ \lambda\gg 1 $) further simplifies to,
\begin{eqnarray}
\Delta -J=\frac{\sqrt{\lambda}}{\pi}|\sin \frac{p}{2}|\label{M}
\end{eqnarray} 
where, $ p $ is the momentum associated with the single magnon excitation.

In the dual string theory description, the magnon dispersion relation (\ref{M}) was recovered by Hofman and Maldacena \cite{Hofman:2006xt} by considering open strings in $ R \times S^{2} $ which is a subspace of the full $ AdS_{5}\times S^{5} $ geometry. In the dual string theory description, these magnon excitation could be thought of as being localized solitonic excitation propagating over an infinitely long string moving in $ R \times S^{2} $. In their analysis \cite{Hofman:2006xt}, Hofman and Maldacena considered open strings rotating on the equator of $ S^{2} $ that maintains a constant angular separation ($ \Delta \varphi $) between its two endpoints which they finally identify as a geometric realization of the magnon momentum ($ p $) associated with the spin chain system. 

It turns out that apart from having the elementary magnon excitation, the asymptotic spectrum associated with the spin chain system also contains bound states of magnon excitation which therefore motivates one to go one step further and generalize the earlier observations \cite{Hofman:2006xt} for multispin magnon excitation, in particular for two spin magnon bound states \cite{Dorey:2006dq}-\cite{Kalousios:2009mp} for which the dispersion relation takes the following form,
\begin{eqnarray}
E-J_{1}=\sqrt{J^{2}_{2}+\frac{\lambda}{\pi^{2}}\sin^{2} \frac{p}{2}}\label{D}
\end{eqnarray}
where, $ J_{2} $ is the second spin of the string which corresponds to the number of magnons in a particular bound state characterized by its energy of excitation ($ E $) and the large spin quantum number $ J_1 $. In other words, from the point of view of (\ref{O}), one could think of two spin magnon excitation as being that of the bound state of $ J_2 $ number of $ Y $ excitation. Clearly, for $ J_2 =1$ one recovers the elementary magnon excitation. In principle, the relation (\ref{D}) is valid for all values of $ J_{2} $ and $ \lambda $. However, in the present analysis, we shall be concerned with the limit, $ \lambda \gg 1 $, $ J_1 \rightarrow \infty$ and $ J_{2}\sim \sqrt{\lambda} $. The most remarkable fact about both the dispersion relations (\ref{M}) and (\ref{D}) is that they exhibit a finite difference between two diverging entities ($ E $ and $ J $) of the theory even at finite t'Hooft coupling ($ \lambda $).

The results corresponding to two spin giant magnon configurations had been subsequently generalized for the three spin case \cite{Spradlin:2006wk},\cite{Ryang:2006yq} where considering the dressing techniques \cite{Spradlin:2006wk}, the three spin giant magnon solutions with three conserved charges $ J_{1}(\rightarrow \infty) $, $ J_2 $ and $ \Psi $ were constructed. These solutions could be thought of as being the superposition of two two spin non interacting bound states of magnons with equal and opposite momenta. At this stage, it is noteworthy to mention that there exists a completely equivalent as well as parallel way of looking at magnon excitation in terms of classical sine-Gordon theory which enable us to look at giant magnon solutions as being that of the solitonic solutions within these special classes of integrable models \cite{Bobev:2006fg}. In fact, this issue had been systematically addressed by constructing the two spin magnon excitation as solitonic solutions within the framework of Complex sine-Gordon theory \cite{Bobev:2006fg}. 

In order to understand the goal of the present analysis, it is first customary to note that all the above discussions regarding the multispin magnon bound states are solely based on the integrable structure of superstring theories in $ AdS_{5}\times S^{5}$ background. However, very recently there have been numerous attempts to extend this vision beyond the usual notion of $ AdS_{5}\times S^{5}$ superstrings and in particular to explore the issue of integrability over deformed geometries namely, by constructing one parameter integrable deformations \cite{Delduc:2013qra}-\cite{Matsumoto:2014ubv} of $ AdS_{5}\times S^{5}$ superstring theories those are not related to the so called T duality transformations \cite{Lunin:2005jy}-\cite{Beisert:2008iq}.

Explorations regarding the integrable structure of superstring theories over these deformed geometries turn out to be an absolutely essential question to be addressed from various perspectives. The first and the foremost issue is related to the interpretation of the dual gauge theory corresponding these deformed geometries. The answer to this question should be certainly non trivial and in fact is quite involved. One of the obvious reasons for this lies over the fact that the original $ SO(2,4)\times SO(6) $ isometry associated with the $ AdS_{5}\times S^{5} $ background gets deformed to its Cartan subgroup $ U(1)^{3}\times U(1)^{3} $. One of the logical steps in this connection would be to explore the dispersion relations associated with these multispin bound states over the $ \kappa $- deformed background by considering their low dimensional analogues. A systematic analysis of which is still lacking in the literature and which is thereby worthy of further investigation. 

In the present analysis, we explore multispin magnon like dispersion relations over the $ \kappa $- deformed $ AdS_{3}\times S^{3} $ background \cite{Hoare:2014pna} which is essentially a $ 6D$ truncated version of the original metric corresponding to the deformed $ AdS_{5}\times S^{5} $  superstring model \cite{Hoare:2014pna}. One of the notable features of this reduced model turns out to be the fact that the $ B $ field of the original $ AdS_{5}\times S^{5} $ superstring model simply vanishes during the $ 6D $ reduction procedure \cite{Hoare:2014pna}. The deformed $ AdS_{3}\times S^{3}$ is also interesting from the point of view of its interpolating structure between the pure $ AdS_{3}\times S^{3} $ near the limit, $ \kappa \rightarrow 0$ and that of the $ dS_{3}\times H^{3}$ in the limit, $ \kappa \rightarrow \infty $ \cite{Hoare:2014pna}.

The organization for the rest of the paper is the following. We start our analysis in Section 2, where we construct spin two bound states over $ \kappa $- deformed $ AdS_{3}\times S^{3} $ (in the limit of the large t'Hooft coupling ($ \lambda \gg 1 $)) and solve corresponding world sheet fluctuations associated with the stringy dynamics on deformed $ S^{3} $. In Section 3, we first compute the energy of excitation ($ E$) as well as the two other angular momenta, $ J_{1} (\rightarrow \infty)$ and $ J_{2} $ exactly in the background deformation ($ \kappa $). Our analysis clearly reveals that the usual dispersion relation \cite{Bobev:2006fg} corresponding to dyonic bound states does not hold in the presence of generic $ \kappa $- deformations. Instead it yields a quite nontrivial relation between different conserved charges of the system. These results further indicate that the conventional spin chain description \cite{Minahan:2002ve} might not hold true for these new class of gauge theories. In the second part of our analysis, we explore the dyonic dispersion relation corresponding to weak background deformations where we observe the emerging dispersion relation of the following type,
\begin{eqnarray}
E -J_{1}-\sqrt{J^{2}_{2}+\frac{\lambda}{\pi^{2}}\sin^{2} \frac{p}{2}}=\mathcal{F}(\kappa^{2})
\end{eqnarray}
where, the R.H.S. is a non trivial function of the deformation parameter ($ \kappa $) which however vanishes smoothly in the limit, $ \kappa \rightarrow 0 $.
In order to complete our discussions on multispin dispersion relations, we perform a similar analysis corresponding to spin three bound state in Section 4, where we mostly focus on small $ \kappa $- deformations and carry out almost identical computations similar to that for the spin two case. Like in the two spin case, considering the conformal gauge, we explore the equations of motion as well as the Virasoro constraints associated with the stringy dynamics over the $ \kappa $- deformed background where we finally push ourselves towards the regime of small deformations ($ 0<\kappa <1 $) and correctly identify the allowed parameter space for the spin three magnon configuration \cite{Ryang:2006yq} in the limit of the large t'Hooft coupling ($ \lambda \gg 1 $). However, the notable difference between the spin three configuration and that of the spin two configuration turns out to be the issue associated with the choice of the static gauge condition. This is due to the fact that unlike the two spin configuration, the string time coordinate associated with the three spin configuration exhibits a non trivial dependence \cite{Ryang:2006yq} on the world-sheet coordinates. Finally, in order to arrive at the desired dispersion relation, we regulate all the UV divergences associated with various conserved entities of the theory. Like in the two spin case, the final dispersion relation corresponding to three spin (giant) configuration also exhibits non trivial $ \kappa $ dependence which we estimate analytically upto quadratic order in the background deformations ($ \kappa $). Finally, we conclude in Section 5.

\section{Two spin magnons: Preliminaries}
We start our analysis by considering the dynamics of open strings on deformed $  AdS_{3}\times S^{3} $ backgrounds that has been recently initiated by the authors in \cite{Delduc:2013qra}-\cite{Hoare:2014pna} and then subsequently explored in many other directions \cite{Khouchen:2014kaa}-\cite{Khouchen:2015jfa},\cite{Banerjee:2015nha}-\cite{Banerjee:2016xbb},\cite{Panigrahi:2014sia}.  The deformed $  AdS_{3}\times S^{3} $ background could be formally expressed as \cite{Hoare:2014pna},
\begin{eqnarray}
ds^{2}&=&ds_{AdS_{3}}^{2}+ds_{S^{3}}^{2}\nonumber\\
ds_{AdS_{3}}^{2}&=&-\mathfrak{h}(\varrho)dt^{2}+\mathfrak{f}(\varrho)d\varrho^{2}+\varrho^{2}d\psi^{2}\nonumber\\
ds_{S^{3}}^{2}&=&\tilde{\mathfrak{h}}(r)d\varphi^{2}+\tilde{\mathfrak{f}}(r)dr^{2}+r^{2}d\phi^{2}
\label{E1}
\end{eqnarray}
where, the metric coefficients turn out to be,
\begin{eqnarray}
\mathfrak{h}&=&\frac{1+\varrho^{2}}{1-\kappa^{2}\varrho^{2}},~~\mathfrak{f}=\frac{1}{(1+\varrho^{2})(1-\kappa^{2}\varrho^{2})}\nonumber\\
\tilde{\mathfrak{h}}&=&\frac{1-r^{2}}{1+\kappa^{2}r^{2}},~~\tilde{\mathfrak{f}}=\frac{1}{(1-r^{2})(1+\kappa^{2}r^{2})}
\end{eqnarray}
such that the NS-NS two form vanishes.

Our goal would be to explore the dispersion relation corresponding to two spin giant magnons on the above background (\ref{E1}) in the limit of large t'Hooft coupling where one of the spin takes large value. We consider two spins of the magnon excitation to be in $ S^{3} $. 

We start our analysis with the Polyakov action for the open string on the deformed background (\ref{E1}),
\begin{eqnarray}
S=-\frac{T}{2}\int_{- \pi}^{\pi}d\sigma d\tau \sqrt{-\gamma}\gamma^{\alpha \beta}\mathfrak{g}_{ab}(X)\partial_{\alpha}X^{a}\partial_{\beta}X^{b}
\label{E3}
\end{eqnarray}
where, the effective string tension could be formally expressed as \cite{Hoare:2014pna},
\begin{eqnarray}
T=\frac{\sqrt{\lambda}}{2 \pi}\sqrt{1+\kappa^{2}}.
\end{eqnarray}

Here, $ \gamma_{\alpha \beta} $ is the induced metric on the world sheet and $ X^{a} $s are the coordinates of the target space. Moreover, here $ \mathfrak{g}_{ab} (X)$ is the metric of the target space. 

The central role of our present discussion is played by the so called Virasoro constraints which are expressed in terms of the components of the stress tensor, 
\begin{eqnarray}
T_{\alpha \beta}=\mathfrak{g}_{ab}\partial_{\alpha}X^{a}\partial_{\beta}X^{b}-\frac{1}{2}\gamma_{\alpha \beta}\gamma^{\mu \nu}\partial_{\mu}X^{a}\partial_{\nu}X^{b}\mathfrak{g}_{ab}=0.
\label{E6}
\end{eqnarray} 

Before we actually proceed further, it is customary to note that the background (\ref{E1}) is invariant under the following translations namely,
\begin{eqnarray}
\delta X^{k}=a^{k},~~k=t,\psi ,\varphi , \phi 
\end{eqnarray}
which therefore suggests that these symmetries could also be realized in the Polyakov action (\ref{E3}). As a result of this, it is indeed quite instructive to write down the conserved charges in the following form,
\begin{eqnarray}
\mathfrak{P}_{k}=T \int_{-\pi}^{\pi} d\sigma \sqrt{-\gamma}\gamma^{\alpha \tau}\mathfrak{g}_{ak}\partial_{\alpha}X^{a}
\end{eqnarray}
where, corresponding to each of these k's one should be able to recover different conserved charges present in the system. For example, the choice, $ k=t $ should give us the energy ($ E $) of the stringy configuration. 

In order to proceed further, we choose the following ansatz,
\begin{eqnarray}
t=\xi \tau ,~~\varrho = \varrho (\tau),~~\psi = \zeta \tau , ~~r=r(\sigma ,\tau),~~\varphi =\varphi (\sigma , \tau),~~\phi = \phi (\sigma , \tau)
\label{E9}
\end{eqnarray}
along with the conformal gauge conditions namely, $ -\gamma^{\tau\tau}=\gamma^{\sigma \sigma}=1 $ and $ \gamma^{\tau \sigma}=0 $ where, $ \xi $ and $ \zeta $ are two constants.

With the above choice (\ref{E9}) in hand,  we first go for some consistency checks. Consider the equation corresponding to, $ X^{t}=t $ which yields,
\begin{eqnarray}
\partial_{\tau}\varrho =0.
\end{eqnarray} 
This clearly suggests that, $ \varrho =\varrho_{0} $ is basically a constant in $ \tau $. In order to fix this constant, we compute the conserved charge corresponding to time translation namely,
\begin{eqnarray}
\mathfrak{P}_{t}=-E=-2 \pi T \xi \mathfrak{h}(\varrho_{0})
\end{eqnarray}
which finally yields,
\begin{eqnarray}
\mathfrak{h}(\varrho_{0})=\frac{E}{2 \pi \xi T}.
\label{E12}
\end{eqnarray}

On the other hand, the equation corresponding to, $ X^{\varrho}=\varrho $ yields,
\begin{eqnarray}
\frac{\xi^{2} \varrho_{0}(1+\kappa^{2})}{(1-\kappa^{2}\varrho_{0}^{2})^{2}}-\varrho_{0}\zeta^{2} =0.
\label{E13}
\end{eqnarray}
Now, this yields two possibilities. One is the possibility that, $ \varrho_{0}=0 $. The other possibility comes through the combination of (\ref{E12}) and (\ref{E13}) which finally yields,
\begin{eqnarray}
\varrho_{0}=\left(\frac{E}{2 \pi T \zeta}\sqrt{1+\kappa^{2}} -1\right)^{1/2}.
\label{E14}
\end{eqnarray}
This further puts constraints on the energy namely,
\begin{eqnarray}
E \geq \frac{2 \pi T \zeta}{\sqrt{1+\kappa^{2}}}.
\end{eqnarray}
Finally, we note that following our construction the equation corresponding to $ \psi $ is trivially satisfied. This further yields the corresponding conserved charge as,
\begin{eqnarray}
\mathfrak{P}_{\psi}=2 \pi T \zeta \varrho_{0}^{2}.
\label{E16}
\end{eqnarray}
In our analysis, however, we consider the first possibility namely, $ \varrho_{0}=0 $ which yields, $ \mathfrak{h}(\varrho_{0})=1 $ and, $ \mathfrak{P}_{\psi}=0 $. With this choice, the effective background geometry seen by the string eventually reduces to, $ R \times S^{3} $ which is a subspace of the full deformed geometry.

Our next task would be to explore the dynamics of the rest of the three other basic variables. In order to do that, we closely follow the method developed in \cite{Arutyunov:2006gs},\cite{Khouchen:2014kaa} where we consider the Virasoro constraints (\ref{E6}) rather than considering the equations of motion directly. It turns out that these Virasoro constraints for the present system yield,
\begin{eqnarray}
T_{\sigma \sigma}=-\frac{E^{2}}{4 \pi^{2}T^{2}}+\tilde{\mathfrak{h}}(r)((\partial_{\sigma}\varphi)^{2}+(\partial_{\tau}\varphi)^{2})+\tilde{\mathfrak{f}}(r)((\partial_{\sigma}r)^{2}+(\partial_{\tau}r)^{2})\nonumber\\
+r^{2}((\partial_{\sigma}\phi)^{2}+(\partial_{\tau}\phi)^{2})=0=T_{\tau \tau}\nonumber\\
T_{\tau \sigma}=\tilde{\mathfrak{h}}(r)\partial_{\sigma}\varphi \partial_{\tau}\varphi +\tilde{\mathfrak{f}}(r)\partial_{\sigma}r \partial_{\tau}r +r^{2}\partial_{\sigma}\phi \partial_{\tau}\phi =0. 
\label{E17}
\end{eqnarray}

The natural next task would be solve these Virasoro constraints. In order to solve these constraints, we choose the following ansatz,
\begin{eqnarray}
\varphi = \omega \tau + \mathfrak{s}(\varsigma),~~\phi = \tau + \mathfrak{q}(\varsigma),~~
r=r(\varsigma)
\label{E18}
\end{eqnarray}
where, the new variable, $ \varsigma = \sigma -\upsilon \omega \tau $ is the linear combination of the world sheet coordinates ($ \sigma, \tau $). Substituting (\ref{E18}) back into (\ref{E17}) we obtain the following set of constraint equations,
\begin{eqnarray}
-\frac{E^{2}}{4 \pi^{2}T^{2}}+\tilde{\mathfrak{h}}(r)(\mathfrak{s}'^{2}+(\omega - \upsilon \omega \mathfrak{s}')^{2})+\tilde{\mathfrak{f}}(r)r'^{2}(1+(\upsilon \omega)^{2})+r^{2}(\mathfrak{q}'^{2}+(1 - \upsilon \omega \mathfrak{q}')^{2})=0\nonumber\\
\tilde{\mathfrak{h}}(r)\mathfrak{s}' \omega (1-\upsilon \mathfrak{s}')-\upsilon \omega\tilde{\mathfrak{f}}(r)r'^{2}+r^{2}\mathfrak{q}'(1-\upsilon\omega \mathfrak{q}')=0\label{eqn23}
\end{eqnarray}
where, the prime indicates derivative w.r.t the variable $ \varsigma $. Clearly, for two spin magnons we encounter a different situation where one needs to solve for three different variables instead of two. This is due to obvious reasons as we have included an additional conserved quantity (charge) into our theory namely, the second angular momentum ($ J_{2} $).

After some trivial algebra we note,
\begin{eqnarray}
\mathfrak{s}'=\frac{\upsilon}{\tilde{\mathfrak{h}}(r)}\frac{(\xi^{2}-\omega^{2}\tilde{\mathfrak{h}}(r))}{(1-\upsilon^{2}\omega^{2})}-\frac{r^{2}\upsilon}{\tilde{\mathfrak{h}}(r)} \frac{(1+\frac{\mathfrak{q}'}{\omega \upsilon}(1-\upsilon^{2} \omega^{2}))}{(1-\upsilon^{2}\omega^{2})}\nonumber\\
r'^{2}=\frac{\Xi (r)}{\tilde{\mathfrak{f}}(r)\tilde{\mathfrak{h}}(r)(1-\omega^{2}\upsilon^{2})^{2}}+\frac{r^{2}\mathfrak{q}'}{\tilde{\mathfrak{f}}(r)\upsilon \omega}(1-\upsilon \omega \mathfrak{q}')\label{eqn24}
\end{eqnarray}
where, the function, $ \Xi (r) $ could be formally expressed as,
\begin{eqnarray}
\Xi (r)=(\xi^{2}-\omega^{2}\tilde{\mathfrak{h}}(r)-r^{2}(1+\frac{\mathfrak{q}'}{\upsilon \omega}(1-\upsilon^{2} \omega^{2})))(\tilde{\mathfrak{h}}(r)-\upsilon^{2}\xi^{2}+\upsilon^{2}r^{2}(1+\frac{\mathfrak{q}'}{\upsilon \omega}(1-\upsilon^{2} \omega^{2}))).
\end{eqnarray}

Our next task would be substitute $ \mathfrak{q}' $ using its E.O.M. which for the present case yields,
\begin{eqnarray}
\mathfrak{q}' \approx -\frac{\upsilon \omega}{(1-\upsilon^{2}\omega^{2})}.
\label{E22}
\end{eqnarray}

Using (\ref{E22}), one finally obtains,
\begin{eqnarray}
\mathfrak{s}'&\approx &\frac{\upsilon}{\tilde{\mathfrak{h}}(r)}\frac{(\xi^{2}-\omega^{2}\tilde{\mathfrak{h}}(r))}{(1-\upsilon^{2}\omega^{2})}\nonumber\\
r'^{2}&\approx &\frac{(\xi^{2}-\omega^{2}\tilde{\mathfrak{h}}(r))(\tilde{\mathfrak{h}}(r)-\upsilon^{2}\xi^{2})-r^{2}\tilde{\mathfrak{h}}(r)}{\tilde{\mathfrak{f}}(r)\tilde{\mathfrak{h}}(r)(1-\omega^{2}\upsilon^{2})^{2}}=-\frac{(r^{2}-r^{2}_{min})(r^{2}_{max}-r^{2})}{(1-r^{2})\tilde{\mathfrak{f}}(r)(1-\omega^{2}\upsilon^{2})^{2}}.
\label{E23}
\end{eqnarray}
where, $ r_{max} $ and $ r_{min} $ correspond to the two extremum values for which the function $ r' $ vanishes. These for the present case turn out to be,
\begin{eqnarray}
r_{min,max}=\sqrt{\frac{1-\alpha (\beta^{2}+\gamma^{2})}{2|1-\alpha|}}\left( 1\pm \sqrt{1+\frac{4\beta^{2}\gamma^{2}\alpha |1-\alpha| }{(1-\alpha (\beta^{2}+\gamma^{2}))^{2}}}\right)^{1/2}
\label{E24}
\end{eqnarray}
where, $ \pm $ correspond to the minimum and the maximum values respectively. Furthermore, different parameters appearing in (\ref{E24}) could be formally expressed as,
\begin{eqnarray}
\beta =\sqrt{\frac{\omega^{2}-\xi^{2}}{\omega^{2}+\kappa^{2}\xi^{2}}},~\gamma = \sqrt{\frac{1-\upsilon^{2}\xi^{2}}{1+\kappa^{2}\upsilon^{2}\xi^{2}}},~\alpha =(\omega^{2}+\kappa^{2}\xi^{2})(1+\kappa^{2}\upsilon^{2}\xi^{2}).
\end{eqnarray}

\section{Dispersion relation}
\subsection{Exact results}
With the above machinery in hand, we now proceed towards computing various conserved charges associated with the stringy dynamics in the bulk. We first compute the conserved charge associated with the angular coordinate $ \varphi $ namely,
\begin{eqnarray}
\mathfrak{P}_{\varphi}=J_{1}=2T \int_{r_{min}}^{r_{max}}\frac{\omega}{|r'|}\tilde{\mathfrak{h}}(r)(\upsilon \mathfrak{s}' -1)dr.
\end{eqnarray}

Using (\ref{E23}), this further yields,
\begin{eqnarray}
J_{1}=2\omega T (1+\kappa^{2}\upsilon^{2}\xi^{2})\int_{r_{min}}^{r_{max}}\frac{(r^{2}-r_{0}^{2})}{(1+\kappa^{2}r^{2})^{3/2}}\frac{dr}{\sqrt{(r^{2}-r^{2}_{min})(r^{2}_{max}-r^{2})}}
\label{E27}
\end{eqnarray}
where, $ r_{0}^{2}=\frac{1-\upsilon^{2}\xi^{2}}{1+\kappa^{2}\upsilon^{2}\xi^{2}} $. 

Our next task would be to perform the above integral (\ref{E27}) and identify the limit where it diverges. We consider the lower bound as, $ r_{min}\rightarrow 0 $ \cite{Khouchen:2014kaa} that turns out to be the correct limit in order to produce states with large angular momentum. Performing the integral (\ref{E27}) in the above limit we find,
\begin{eqnarray}
J_{1}=2\omega T \mathcal{I}(r)|_{r=\epsilon}^{r=r_{max}}\label{E28}
\end{eqnarray}
where, the \textit{exact} analytic form corresponding to the function $ \mathcal{I}(r) $ could be formally expressed as,
\begin{eqnarray}
\mathcal{I}(r)=\frac{\left(\kappa ^2 \xi ^2 \upsilon ^2+1\right)\mathcal{N}(r)}{\mathcal{D}(r)}
\end{eqnarray}
where, the explicit form of the functions are as follows,
\begin{eqnarray}
\mathcal{N}(r)= r_0^2 r \left(\kappa ^2 r_{max}^2+1\right) \sqrt{ \frac{r_{max}^2}{r^2}-1} \sqrt{\frac{1}{\kappa ^2 r^2}+1} F_1\left(1;\frac{1}{2},\frac{1}{2};2;\frac{r_{max}^2}{r^2},-\frac{1}{r^2 \kappa ^2}\right)\nonumber\\
+2 r \left(\kappa ^2r_0^2+1\right) \left(r^2-r_{max}^2\right)
\end{eqnarray}
\begin{eqnarray}
\mathcal{D}(r)=2 \left(\kappa ^2 r_{max}^2+1\right) \sqrt{r^2 \left(r_{max}^2-r^2\right)} \sqrt{\kappa ^2 r^2+1}
\end{eqnarray}
such that $ F_{1} $ is the so called Appell function of the first kind and $ |\epsilon | \ll 1 $ is some suitable cutoff for the theory.

Next, we compute the energy ($ E =2\pi \xi T$) associated with the stringy configuration corresponding to this large $ J_1 $ limit. In order to do that, we first note,
\begin{eqnarray}
2 \pi &=& \int_{- \pi}^{\pi} d\sigma = 2\int_{0}^{r_{max}}\frac{dr}{|r'|}=\mathfrak{K}(r_{max})-\mathfrak{K}(\epsilon)\label{Ee36}
\end{eqnarray}
where, the entity, $ \mathfrak{K}(r) $ could be formally expressed as,
\begin{eqnarray}
\mathfrak{K}(r)=\frac{r \sqrt{\frac{r^{2}_{max}}{r^2}-1} \left(\upsilon ^2 \omega ^2-1\right) \sqrt{\frac{1}{\kappa ^2 r^2}+1} F_1\left(1;\frac{1}{2},\frac{1}{2};2;\frac{r^{2}_{max}}{r^2},-\frac{1}{r^2 \kappa ^2}\right)}{2 \sqrt{r^2 (r^{2}_{max}-r^{2})} \sqrt{\kappa ^2 r^2+1}}
\end{eqnarray}
which is again an \textit{exact} expression in the background deformations ($ \kappa $). 

The corresponding energy turns out to be,
\begin{eqnarray}
E=\xi T(\mathfrak{K}(r_{max})-\mathfrak{K}(\epsilon))
\end{eqnarray}
which finally yields,
\begin{eqnarray}
E-J_1 =\frac{T\mathcal{Z}(r)|_{\epsilon}^{r_{max}}}{2 \sqrt{r^2 (r^{2}_{max}-r^{2})} \sqrt{\kappa ^2 r^2+1}}\label{e38}
\end{eqnarray}
where, the entity $ \mathcal{Z}(r) $ could be formally expressed as,
\begin{eqnarray}
\mathcal{Z}(r)=-r \left(\sqrt{\frac{r^{2}_{max}}{r^2}-1}\right) \sqrt{\frac{1}{\kappa ^2 r^2}+1}~ F_1\left(1;\frac{1}{2},\frac{1}{2};2;\frac{r^{2}_{max}}{r^2},-\frac{1}{r^2 \kappa ^2}\right)\nonumber\\ \left(-\xi  \upsilon ^2 \omega ^2+\xi +2 r_{0}^2 \omega  \left(\kappa ^2 \xi ^2 \upsilon ^2+1\right)\right)+\frac{4 r \omega  \left(\kappa ^2 r_{0}^2+1\right) (r^{2}_{max}-r^{2})\left(\kappa ^2 \xi ^2 \upsilon ^2+1\right)}{\kappa ^2 r_{max}^2+1}.
\end{eqnarray}

Finally, we are left with the third and the last conserved charge associated with our system namely, the second angular momentum, $ J_{2} $ which for the present case turns out to be,
\begin{eqnarray}
\mathfrak{P}_{\phi}=J_{2}=-T\int_{-\pi}^{\pi} d\sigma r^{2}\partial_{\tau}\phi =\frac{2 T\tan ^{-1}(\kappa r_{max})}{\kappa }\label{e37}
\end{eqnarray}
which could be subsequently inverted in order to express,
\begin{eqnarray}
r_{max}=\frac{1}{\kappa}\tan \left(\frac{\kappa J_{2}}{2T} \right)=\frac{1}{\kappa}\tan\mathfrak{j} \kappa\label{e41} 
\end{eqnarray}
where, we have set the new parameter, $ \mathfrak{j} = \frac{J_2}{2T} $.

Substituting, (\ref{e41}) into (\ref{e38}) we find,
\begin{eqnarray}
E-J_1 =\frac{\kappa T\mathcal{Z}(r)|_{\epsilon}^{r_{max}}}{2 \sqrt{r^2 (\tan^{2}\mathfrak{j}\kappa -\kappa^{2}r^{2})} \sqrt{\kappa ^2 r^2+1}}\label{eq42}
\end{eqnarray}
where, we replace $ r_{max} $ in order to express,
\begin{eqnarray}
\mathcal{Z}(r)=-r \left(\sqrt{\frac{\tan^{2}\mathfrak{j} \kappa}{\kappa^{2}r^2}-1}\right) \sqrt{\frac{1}{\kappa ^2 r^2}+1}~ F_1\left(1;\frac{1}{2},\frac{1}{2};2;\frac{\tan^{2}\mathfrak{j} \kappa}{\kappa^{2}r^2},-\frac{1}{r^2 \kappa ^2}\right)\nonumber\\ \left(-\xi  \upsilon ^2 \omega ^2+\xi +2 r_{0}^2 \omega  \left(\kappa ^2 \xi ^2 \upsilon ^2+1\right)\right)+\frac{4 r \omega  \left(\kappa ^2 r_{0}^2+1\right) (\tan^{2}\mathfrak{j}\kappa -\kappa^{2}r^{2})\left(\kappa ^2 \xi ^2 \upsilon ^2+1\right)}{\kappa^{2}\sec^{2}\mathfrak{j}\kappa }.
\end{eqnarray}

The above equation (\ref{eq42}) might be regarded as the new dispersion relation corresponding to spin two bound states over $ \kappa $- deformed background that relates various conserved charges associated with the bound state. However, in order to explore whether the R.H.S of (\ref{eq42}) contains any explicit momentum dependence as that is observed for usual dyonic systems \cite{Chen:2006gea}, one needs to compute the momentum ($ p $) associated with these two spin excitation. In AdS/CFT, the magnon momentum ($ p $) associated with the one dimensional spin chain could be realized as a geometric entity that is basically related to the angle of separation ($ \Delta \varphi $) between the two end points of the string \cite{Hofman:2006xt},
\begin{eqnarray}
\frac{p}{2} &=&\frac{\Delta \varphi}{2} =\int_{0}^{r_{max}}\frac{\varphi'}{|r'|}dr\nonumber\\
&\approx &\int_{0}^{r_{max}}\frac{r (1+\kappa^{2}\upsilon^{2}\xi^{2})}{\upsilon(1-r^{2})\sqrt{r_{max}^{2}-r^{2}}}\frac{dr}{\sqrt{1+\kappa^{2}r^{2}}}
\end{eqnarray}
where, we have used the fact that, $ |r^{2}_{0}/r^{2}| \ll 1 $.

A straightforward computation yields a rather non trivial answer,
\begin{eqnarray}
\Delta \varphi =p =\frac{\kappa(\kappa ^2 \xi ^2 \upsilon ^2+1)\mathcal{A}(\kappa)}{\upsilon  \sqrt{\kappa ^2+1} \sqrt{\tan^{2}\mathfrak{j}\kappa -\kappa^{2}} }\label{e42}
\end{eqnarray}
where, the entity, $ \mathcal{A}(\kappa) $ above in (\ref{e42}) could be formally expressed as,
\begin{eqnarray}
\mathcal{A}(\kappa)=\log \left(\frac{\tan^{2}\mathfrak{j}\kappa}{\kappa^{2}}-1\right)-\log \left(\left(1-\frac{\tan^{2}\mathfrak{j}\kappa}{\kappa^{2}}\right) \left(\tan^{2}\mathfrak{j}\kappa +1\right)\right)\nonumber\\
+\log \left(-\left(\kappa ^2+2\right) \frac{\tan^{2}\mathfrak{j}\kappa}{\kappa^{2}}-2 \sqrt{\kappa ^2+1} \sqrt{\frac{\tan^{2}\mathfrak{j}\kappa}{\kappa^{2}}} \sqrt{\frac{\tan^{2}\mathfrak{j}\kappa}{\kappa^{2}}-1}+1\right).\label{eq46}
\end{eqnarray}

Using (\ref{eq46}) we further note,
\begin{eqnarray}
p=\frac{\kappa(\kappa ^2 \xi ^2 \upsilon ^2+1)}{\upsilon  \sqrt{\kappa ^2+1} \sqrt{\tan^{2}\mathfrak{j}\kappa -\kappa^{2}} }\log \left( \frac{\Omega (\mathfrak{j}\kappa)}{\sec^{2}\mathfrak{j}\kappa}\right)\label{eq47} 
\end{eqnarray}
where,
\begin{eqnarray}
\Omega (\mathfrak{j}\kappa)=\left(\kappa ^2+2\right) \frac{\tan^{2}\mathfrak{j}\kappa}{\kappa^{2}}+2 \sqrt{\kappa ^2+1} \sqrt{\frac{\tan^{2}\mathfrak{j}\kappa}{\kappa^{2}}} \sqrt{\frac{\tan^{2}\mathfrak{j}\kappa}{\kappa^{2}}-1}-1.\label{omega}
\end{eqnarray}

The above equation (\ref{eq47}) could be inverted in order to express the R.H.S. of (\ref{eq42}) as,
\begin{eqnarray}
E-J_1 =\frac{\kappa T\mathcal{Z}(r, p)|_{\epsilon}^{r_{max}}}{2 \sqrt{r^2 (\tan^{2}\mathfrak{j}\kappa -\kappa^{2}r^{2})} \sqrt{\kappa ^2 r^2+1}}\label{eq49}
\end{eqnarray}
where, the function $ \mathcal{Z}(r, p) $ could be formally expressed as,
\begin{eqnarray}
\mathcal{Z}(r, p)=-r \left(\sqrt{\frac{\tan^{2}\mathfrak{j} \kappa}{\kappa^{2}r^2}-1}\right) \sqrt{\frac{1}{\kappa ^2 r^2}+1}~ F_1\left(1;\frac{1}{2},\frac{1}{2};2;\frac{\tan^{2}\mathfrak{j} \kappa}{\kappa^{2}r^2},-\frac{1}{r^2 \kappa ^2}\right)\nonumber\\ \left(-\xi  \upsilon ^2 \omega ^2+\xi +2 r_{0}^2 \omega  \left(\kappa ^2 \xi ^2 \upsilon ^2+1\right)\right)+\frac{4 r \omega  \left(\kappa ^2 r_{0}^2+1\right) (\tan^{2}\mathfrak{j}\kappa -\kappa^{2}r^{2})\left(\kappa ^2 \xi ^2 \upsilon ^2+1\right)}{\kappa^{2}\Omega (\mathfrak{j}\kappa)e^{-\chi p} }
\end{eqnarray}
together with the function,
\begin{eqnarray}
\chi =\frac{\upsilon  \sqrt{\kappa ^2+1} \sqrt{\tan^{2}\mathfrak{j}\kappa -\kappa^{2}}}{\kappa(\kappa ^2 \xi ^2 \upsilon ^2+1)}.\label{cai}
\end{eqnarray}

Substituting the limits, we finally obtain,
\begin{eqnarray}
\frac{E-J_{1}}{\kappa T}=-\frac{1}{2\kappa^{2}r^{2}_{max}}F_1\left(1;\frac{1}{2},\frac{1}{2};2;\frac{\tan^{2}\mathfrak{j} \kappa}{\kappa^{2}r_{max}^2},-\frac{1}{r_{max}^2 \kappa ^2}\right)\left(-\xi  \upsilon ^2 \omega ^2+\xi +2 r_{0}^2 \omega  \left(\kappa ^2 \xi ^2 \upsilon ^2+1\right)\right)\nonumber\\+\frac{2 \omega  \left(\kappa ^2 r_{0}^2+1\right) (\tan^{2}\mathfrak{j}\kappa -\kappa^{2}r_{max}^{2})^{1/2}\left(\kappa ^2 \xi ^2 \upsilon ^2+1\right)}{\kappa^{2}\Omega (\mathfrak{j}\kappa)e^{-\chi p}\sqrt{1+\kappa^{2}r^{2}_{max}} }-\frac{2 \omega  \left(\kappa ^2 r_{0}^2+1\right) (\tan^{2}\mathfrak{j}\kappa )^{1/2}\left(\kappa ^2 \xi ^2 \upsilon ^2+1\right)}{\kappa^{2}\Omega (\mathfrak{j}\kappa)e^{-\chi p} }\nonumber\\+\frac{1}{2\kappa^{2}\epsilon^{2}}F_1\left(1;\frac{1}{2},\frac{1}{2};2;\frac{\tan^{2}\mathfrak{j} \kappa}{\kappa^{2}\epsilon^{2}},-\frac{1}{ \kappa ^2 \epsilon^{2}}\right)\left(-\xi  \upsilon ^2 \omega ^2+\xi +2 r_{0}^2 \omega  \left(\kappa ^2 \xi ^2 \upsilon ^2+1\right)\right).
\end{eqnarray}

If we now demand that the finite dispersion relation is finite (in the sense that it is free from $ \epsilon $ divergences) then this amounts of setting the following constraint between various parameters of the theory namely,
\begin{eqnarray}
-\xi  \upsilon ^2 \omega ^2+\xi +2 r_{0}^2 \omega  \left(\kappa ^2 \xi ^2 \upsilon ^2+1\right)=0
\end{eqnarray}
which finally yields,
\begin{eqnarray}
\frac{E-J_{1}}{ T}=-\frac{2 \omega  \left(\kappa ^2 r_{0}^2+1\right) (\tan^{2}\mathfrak{j}\kappa )^{1/2}\left(\kappa ^2 \xi ^2 \upsilon ^2+1\right)}{\kappa \Omega (\mathfrak{j}\kappa)e^{-\chi p} }.\label{e55}
\end{eqnarray}

The above relation (\ref{e55}) is the final version of the dispersion relation including both the momentum ($ p $) as well as the second conserved charge ($ J_{2} $). Clearly, the usual dispersion relation \cite{Chen:2006gea} corresponding to dyonic bound states does not hold for generic $ \kappa $- deformations \cite{Hoare:2014pna}. These results therefore strongly suggest that in the limit of the large t'Hooft coupling ($ \lambda \gg 1 $), the corresponding excitation in the dual gauge theory cannot be interpreted as magnons in the usual sense. In the next section we would explore the corresponding deviation in the limit of weak deformations which should correctly reproduce the known dispersion relation in the limit $ \kappa \rightarrow 0 $.

Finally, as a consistency check of our analysis, we consider the limit, $ J_2 \rightarrow 0 $ in (\ref{e37}) which for the present case could be realized by setting, $ \phi =C $ where, $ C $ is some constant \cite{Khouchen:2014kaa}. In other words, we excite only one of the angular modes associated to $ S^{3} $. As a consequence of this using Virasoro constraints (\ref{E17}) we find,
\begin{eqnarray}
-\xi^{2}+\tilde{\mathfrak{h}}(r)(\mathfrak{s}'^{2}+(\omega - \upsilon \omega \mathfrak{s}')^{2})+\tilde{\mathfrak{f}}(r)r'^{2}(1+(\upsilon \omega)^{2})=0\nonumber\\
\tilde{\mathfrak{h}}(r)\mathfrak{s}' \omega (1-\upsilon \mathfrak{s}')-\upsilon \omega\tilde{\mathfrak{f}}(r)r'^{2}=0
\end{eqnarray}
which could be solved quite easily in order to obtain,
\begin{eqnarray}
\mathfrak{s}'&=&\frac{\upsilon}{\tilde{\mathfrak{h}}(r)}\frac{(\xi^{2}-\omega^{2}\tilde{\mathfrak{h}}(r))}{(1-\upsilon^{2}\omega^{2})}\nonumber\\
r'^{2}&=&\frac{\omega^{2}(\frac{\xi^{2}}{\omega^{2}}-\tilde{\mathfrak{h}}(r))(\tilde{\mathfrak{h}}(r)-\upsilon^{2}\xi^{2})}{\tilde{\mathfrak{f}}(r)\tilde{\mathfrak{h}}(r)(1-\omega^{2}\upsilon^{2})^{2}}
\end{eqnarray}
that precisely matches with the earlier findings \cite{Khouchen:2014kaa} in the context of giant magnon dispersion relation with one spin. As a consequence of this, the corresponding dispersion relation \cite{Khouchen:2014kaa} for gaint magnons on deformed $ AdS_3 \times S^{3}$ follows trivially.

\subsection{Perturbative results}
In order to explore the dispersion relation for small background deformations, we first expand the function, $ \mathcal{I}(r) $ perturbatively in $ \kappa $ which yields the following,
\begin{eqnarray}
\mathcal{I}(r)=F_1\left(1;\frac{1}{2},\frac{1}{2};2;\frac{r_{max}^2}{r^2},-\frac{1}{\kappa ^2 r^2}\right)\left(\frac{r_{0}^{2}}{2\kappa r^{2}} \right)(1+\kappa^{2}\xi^{2}\upsilon^{2})\nonumber\\-\frac{\sqrt{r_{max}^{2}-r^{2}}}{2}\left( 2-\kappa ^2 \left(-2 \xi ^2 \upsilon ^2-2 r_0^2+2 r_{max}^2+r^2\right)\right)+\mathcal{O}(\kappa^{3}).\label{Ee47} 
\end{eqnarray}
 Clearly, the above result (\ref{Ee47}) possesses a divergent piece near the limit, $ r \rightarrow 0 $. This essentially corresponds to the large spin ($ J_{1}\rightarrow \infty $) limit of the usual magnon solution \cite{Hofman:2006xt},\cite{Khouchen:2014kaa} as mentioned in the previous Section. We separate the singular piece from the finite contribution which finally yields,
\begin{eqnarray}
J_{1}=2F_1\left(1;\frac{1}{2},\frac{1}{2};2;1,-\frac{1}{\kappa ^2 r_{max}^2}\right)\left( \frac{\omega  T r_{0}^{2}}{2\kappa r_{max}^{2}} \right)(1+\kappa^{2}\xi^{2}\upsilon^{2})\nonumber\\ +2\omega T r_{max}(1+\kappa ^2\left(\xi ^2 \upsilon ^2+r_0^2-r_{max}^{2}\right))-2\omega TJ^{(\epsilon)}_{1}\label{E30}
\end{eqnarray}
where, $ |\epsilon|\ll 1 $ is some cutoff such that the entity,
\begin{eqnarray}
J^{(\epsilon)}_{1}=F_1\left(1;\frac{1}{2},\frac{1}{2};2;\frac{r_{max}^2}{\epsilon ^2},-\frac{1}{\kappa ^2 \epsilon ^2}\right)\left( \frac{ \left(1- \xi ^2 \upsilon ^2\right)}{2 \kappa  \epsilon^2}\right) 
\end{eqnarray}
represents the divergent piece in the angular momentum.

 Next, we expand (\ref{Ee36}) perturbatively in the background deformations ($ \kappa $) which finally yields,
\begin{eqnarray}
2 \pi \approx-2\left( \frac{(1-\omega^{2}\upsilon^{2})}{2 \kappa r^{2}_{max}}\right) F_1\left(1;\frac{1}{2},\frac{1}{2};2;1,-\frac{1}{\kappa ^2 r_{max}^2}\right)+E^{(div)}\label{eq55}
\end{eqnarray}
where,
\begin{eqnarray}
E^{(div)} =2 F_1\left(1;\frac{1}{2},\frac{1}{2};2;\frac{r_{max}^2}{\epsilon ^2},-\frac{1}{\kappa ^2 \epsilon ^2}\right)\left( \frac{(1-\omega^{2}\upsilon^{2})}{2 \kappa \epsilon^{2}}\right)
\end{eqnarray}
possesses the usual divergences. Finally, using (\ref{E30}) and (\ref{eq55}) and setting, $ \omega =-\xi $ we get rid of these divergences and arrive at the following finite entity,
\begin{eqnarray}
E -J_{1}= 2\xi T r_{max}(1+\kappa ^2\left( \xi^{2}\upsilon ^2+r_0^2-r_{max}^{2}\right)).\label{E34}
\end{eqnarray}

Next, we expand the other angular momentum ($ J_{2} $) which for the present case turns out to be,
\begin{eqnarray}
J_{2}\approx 2T r_{max}\left( 1-\frac{\kappa^{2}r^{2}_{max}}{3}\right) +\mathcal{O}(\kappa^{3}).\label{E36}
\end{eqnarray}

Our final task would be to compute the momentum ($ p $) associated with these two spin bound states which for the present case yields, 
\begin{eqnarray}
\frac{p}{2}=\frac{\Delta \varphi}{2} 
\approx \frac{1}{\upsilon}\int_{0}^{r_{max}}\frac{r}{(1-r^{2})}\frac{dr}{\sqrt{r^{2}_{max}-r^{2}}}-\frac{\kappa^{2}}{2\upsilon}\int_{0}^{r_{max}}\frac{r(r^{2}-2\xi^{2}\upsilon^{2})}{(1-r^{2})}\frac{dr}{\sqrt{r^{2}_{max}-r^{2}}}.\label{E37}
\end{eqnarray}

Performing the above integral (\ref{E37}) and rescaling the momentum we finally obtain,
\begin{eqnarray}
\sin \frac{\tilde{p}}{2} &\approx & \mathfrak{j}-  \frac{\kappa^{2}\mathfrak{b}}{2\sqrt{|1-\mathfrak{j}^{2}|}}+\mathcal{O}(\kappa^{6})\nonumber\\
\mathfrak{b}&=&\frac{(2\xi^{2}\upsilon^{2}-1)}{\sqrt{|1-\mathfrak{j}^{2}|}}\sin^{-1}\mathfrak{j}-\mathfrak{j}
\label{E38} 
\end{eqnarray} 
where, we have used the fact that, $ |r^{2}_{0}/r^{2}| \ll 1 $. 

Finally, using (\ref{E34}), (\ref{E36}) and (\ref{E38}) we arrive at the following dispersion relation corresponding to two spin bound states, 
\begin{eqnarray}
E-J_{1}-\sqrt{J^{2}_{2}+\frac{\hat{\lambda}}{\pi^{2}}\sin^{2}\frac{\tilde{p}}{2}}=\mathcal{F}(\kappa^{2})=\sqrt{\lambda}\kappa^{2}\left( \frac{\sqrt{2}\mathfrak{j}}{\pi}\right) \mathcal{K}+\mathcal{O}(\kappa^{4})\label{E39}
\end{eqnarray}
where, $ \hat{\lambda}=\lambda (1+\kappa^{2}) $ and we have set, $ \xi =\sqrt{2} $. The entity, $ \mathcal{K} $ is basically a constant and could be formally expressed as,
\begin{eqnarray}
\mathcal{K}=2\upsilon^{2}+r^{2}_{0}-\frac{1}{2}-\frac{5}{6}\mathfrak{j}^{2}+\frac{ \mathfrak{b}}{4\mathfrak{j}\sqrt{|1-\mathfrak{j}^{2}|}}
\label{E40}
\end{eqnarray}
such that, one could finally replace $ r_{max}|_{\kappa \rightarrow 0} \approx \mathfrak{j}$ in terms of the conserved charge $ J_{2} $ (\ref{E36}) as in the previous analysis.

The above relation (\ref{E39}) essentially summarizes the dispersion formula for small background deformations. In the limit of the vanishing deformation ($ \kappa \rightarrow 0 $), the above expression (\ref{E39}) smoothly reduces to the known dispersion relation for giant magnon bound states corresponding to string sigma models in $ R \times S^{3} $  \cite{Bobev:2006fg}. In summary, from the above analysis, it turns out that in the presence of $ \kappa $ deformations \cite{Hoare:2014pna} the R.H.S. of the dispersion relation (\ref{E39}) corresponding to the two spin bound states could in principle be non vanishing depending on the various parameters of the theory. However, the corresponding interpretation from the perspective of the dual gauge theory is not very clear at this moment and which is thereby worthy of further investigations.
\section{A note on three spin systems}
As a possible extension of our previous analysis, we now study the dispersion relation corresponding to three spin giant magnons over the $ \kappa $- deformed background (\ref{E1}) in the limit of the large t'Hooft ($ \lambda \gg 1 $) coupling. Unlike the previous example, here we mostly focus on the perturbative regime in order to gain an overview of the full theory with background deformations.  We consider one of the spins $ \Psi $ in $ AdS_{3} $ and the other two spins, $ J_{1} $ and $ J_{2} $ along the two orthogonal directions of the three sphere ($ S^{3} $). In order to proceed further, we consider the following ansatz \cite{Ryang:2006yq},
\begin{eqnarray}
t&=& \tau +\mathfrak{t}(\varsigma),~~\varrho=\varrho (\varsigma),~~\varphi = \omega_{1}\tau +\mathfrak{s}(\varsigma)\nonumber\\
\phi &=&  \tau + \mathfrak{q}(\varsigma),~~r=r(\varsigma),~~\psi =\omega_{2}\tau +\mathfrak{p}(\varsigma)
\end{eqnarray}
where, $ \varsigma = \sigma -\upsilon \tau $.

In the following, we first enumerate all the conserved quantities those are of interest to us namely,
\begin{eqnarray}
E &=& T \int_{- \pi}^{\pi}d\sigma \mathfrak{h}(\varrho)(1-\upsilon \mathfrak{t}')\nonumber\\
\Psi &=& -T \int_{- \pi}^{\pi}d\sigma  \varrho^{2}(\omega_{2}-\upsilon \mathfrak{p}')\nonumber\\
J_{1}&=&-T \int_{- \pi}^{\pi}d\sigma \tilde{\mathfrak{h}}(r)(\omega_{1}-\upsilon \mathfrak{s}')\nonumber\\
 J_{2}&=&-T \int_{- \pi}^{\pi}d\sigma r^{2}(1- \upsilon \mathfrak{q}').\label{E41}
\end{eqnarray}

Next, we compute the Virasoro constraints for the present example that could be formally expressed as the following set of equations namely,
\begin{eqnarray}
T_{\tau \tau}=T_{\sigma \sigma}=- \mathfrak{h}(\mathfrak{t}'^{2}+(1-\upsilon \mathfrak{t}')^{2})+\varrho'^{2}f(\varrho)(1+\upsilon^{2})+\varrho^{2}(\mathfrak{p}'^{2}+(\omega_{2}-\upsilon \mathfrak{p}')^{2})\nonumber\\
+\tilde{\mathfrak{h}}(\mathfrak{s}'^{2}+(\omega_{1}-\upsilon \mathfrak{s}')^{2})+r'^{2}\tilde{f}(r)(1+\upsilon^{2})+r^{2}(\mathfrak{q}'^{2}+(1-\upsilon \mathfrak{q}')^{2})=0
\end{eqnarray}
\begin{eqnarray}
T_{\tau \sigma}=- \mathfrak{h}\mathfrak{t}'(1-\upsilon \mathfrak{t}')- \upsilon\varrho'^{2}f(\varrho)+\varrho^{2}\mathfrak{p}'(\omega_{2}-\upsilon \mathfrak{p}')\nonumber\\
+\tilde{\mathfrak{h}}\mathfrak{s}'(\omega_{1}-\upsilon \mathfrak{s}')-\upsilon r'^{2}\tilde{f}(r)+r^{2}\mathfrak{q}'(1-\upsilon \mathfrak{q}')=0.
\end{eqnarray}

After some trivial algebra we find,
\begin{eqnarray}
r'^{2}&=& -\frac{\mathfrak{h}\mathfrak{t}'}{\upsilon \tilde{f}(r)}(1-\upsilon \mathfrak{t}')-\frac{\varrho'^{2}f(\varrho)}{\tilde{f}(r)}+\frac{\varrho^{2}\mathfrak{p}'}{\upsilon \tilde{f}(r)}(\omega_{2}-\upsilon \mathfrak{p}')++\frac{\tilde{\mathfrak{h}}\mathfrak{s}'}{\upsilon \tilde{f}(r)}(\omega_{1}-\upsilon \mathfrak{s}')+\frac{r^{2}\mathfrak{q}'}{\upsilon \tilde{f}(r)}(1-\upsilon \mathfrak{q}').\nonumber\\
\mathfrak{s}'&=&\frac{1}{\tilde{\mathfrak{h}}(r)\omega_{1}}(\mathfrak{h}\mathfrak{t}' -\varrho^{2}\omega_{2}\mathfrak{p}'-r^{2}\mathfrak{q}')+\frac{\upsilon}{\tilde{\mathfrak{h}}(r)\omega_{1}(1-\upsilon^{2})}(\mathfrak{h}-\tilde{\mathfrak{h}}\omega_{1}^{2}-\varrho^{2}\omega_{2}^{2}-r^{2}).\label{E44}
\end{eqnarray}

Our next task would be to solve for these fluctuations using their E.O.Ms (in the limit of the large t'Hooft coupling ($ \lambda \gg 1 $)) and substitute it back into (\ref{E44}). A straightforward computation finally yields,
\begin{eqnarray}
r'^{2}&=&\frac{(1-r^{2})(r^{2}-r^{2}_{min})(r^{2}_{max}-r^{2})}{(1-\upsilon^{2})^{2}}\nonumber\\
\mathfrak{s}'& = &-\frac{\omega_{1}\upsilon}{(1-\upsilon^{2})}
\end{eqnarray}
where, the roots could be formally expressed as,
\begin{eqnarray}
r_{max,min}=\frac{1}{\sqrt{2}\kappa}(\omega_{1}-1+\kappa^{2}(\mathfrak{h}_{c}-\omega_{2}\varrho^{2}_{c}))^{1/2} \left( 1 \pm \sqrt{1-\frac{4\kappa^{2}(\omega_{2}\varrho^{2}_{c}+\omega_{1}-\mathfrak{h}_{c})}{(\omega_{1}-1+\kappa^{2}(\mathfrak{h}_{c}-\omega_{2}\varrho^{2}_{c}))^{2}}}\right)^{1/2}\label{E47}
\end{eqnarray}
subjected to the fact that we have set, $ \mathfrak{h}_{c}=\mathfrak{h}(\varrho =\varrho_{c}) $.
With the above solutions in hand, we now proceed towards evaluating all the conserved quantities in (\ref{E41}). We first compute the energy which turns out to be,
\begin{eqnarray}
E=2 T\mathfrak{h}_{c}\int_{\epsilon_{IR}}^{r_{max}}\frac{dr}{r\sqrt{1-r^{2}}}\frac{1}{\sqrt{r^{2}_{max}-r^{2}}}\label{E48}
\end{eqnarray}
where, like in the previous example, we set the lower limit ($r_{min}$) equal to some IR cutoff, $ \epsilon_{IR} $ which is sufficiently small compared to unity namely, $ |\epsilon_{IR}|\ll 1 $. At this point, the reader should take a note on the fact that this is indeed a valid limit when one of the conserved charges ($ J_{1} \rightarrow \infty $) goes to infinity \cite{Khouchen:2014kaa}. On the other hand, for the sake of convenience we set the upper limit, $ r_{max}=1 $. In the following, we would first like to explore the consequences of setting this upper limit equal to unity. Using, (\ref{E47}) one could try to have an estimate on the deformation parameter ($ \kappa $) for the theory and this simply implies,
\begin{eqnarray}
\kappa^{2}\approx \frac{(1-\omega_{1})^{2}}{(\omega_{1}+2)(2\omega_{1}-1)}\label{49}
\end{eqnarray}
subjected to the choice, $ \omega_{2}\varrho^{2}_{c}=1 $ and, $ \mathfrak{h}_{c}=-\omega_{1} $ which will be considered throughout the subsequent analysis. If we now demand, $ 0<\kappa^{2}<1 $ which in turn implies that we push ourselves towards the limit without any deformation \cite{Ryang:2006yq}, then from (\ref{49}) it immediately follows,
\begin{eqnarray}
\frac{1}{2}<\omega_{1}\leq 1\label{50}
\end{eqnarray}
which thereby non trivially constraints one of the important parameters of the theory which will eventually show up in the final dispersion relation that we are after. Furthermore, (\ref{50}) also confirms that, $ (1-\omega^{2}_{1})\geq 0$ which is also quite consistent with the earlier observations made in connection to that of the three spin systems \cite{Ryang:2006yq}.

Considering these facts and evaluating the integral (\ref{E48}) we find,
\begin{eqnarray}
 E=2T \mathfrak{h}_{c}\log\left( \frac{r}{\sqrt{1-r^{2}}}\right)\vert_{\epsilon_{IR}}^{1}.\label{E49}
\end{eqnarray}

Next, we compute the angular momentum,
\begin{eqnarray}
 J_{1}&=&-2T\omega_{1}\int_{\epsilon_{IR}}^{1}\frac{dr}{r(1+\kappa^{2}r^{2})}\nonumber\\
 &=& -2 T\omega_{1} \log \left( \frac{r}{\sqrt{1+ \kappa^{2}r^{2}}}\right)\nonumber\\ 
 &= &-2\omega_{1}T \log r|_{\epsilon_{IR}}^{1}+ \kappa^{2}\omega_{1}T +\mathcal{O}(\kappa^{3}).\label{E50}
\end{eqnarray}

Using, (\ref{E49}) and (\ref{E50}) we finally note the difference,
\begin{eqnarray}
E-J_{1}=-\frac{\sqrt{\lambda}\kappa^{2}\omega_{1}}{2 \pi}\sqrt{1+\kappa^{2}}+\mathcal{A}^{(div)}\label{E50}
\end{eqnarray}
where, we separate out the divergent piece as,
\begin{eqnarray}
\mathcal{A}^{(div)}=(\mathfrak{h}_{c}+\omega_{1})\log r|_{\epsilon_{IR}}^{1}-\frac{\mathfrak{h}_{c}}{2}\log (1-r^{2})|_{\epsilon_{IR}}^{1}.\label{E51}
\end{eqnarray}

Therefore, like in the previous example corresponding to two spin bound systems, we note that both $ E $ and $ J_{1} $ as well as their difference corresponding to the three spin configuration diverges as well. From (\ref{E51}), we note that the difference, $ E-J_{1} $ possesses both UV as well as IR divergences. However, by setting, $ \mathfrak{h}_{c}=-\omega_{1} $ one could in principle get rid off the IR divergences which thereby precisely falls in agreement with the earlier observations in the context of three spin giant magnon configurations \cite{Ryang:2006yq}.

We now go for computing the following conserved quantity,
\begin{eqnarray}
 \Psi= -2T \int_{\epsilon_{IR}}^{1}\frac{dr}{r(1-r^{2})}\label{E53}
\end{eqnarray}
where, we set, $ \varrho^{2}_{c}\omega_{2}=1 $. Clearly, the above integral (\ref{E53}) is divergent both at UV as well as near the IR limit. For the moment, we may forget about the IR divergence and try to regularize the divergence near the UV limit ($ r\rightarrow 1$). In order to do that, we set an UV cutoff, $  |\Lambda|\lesssim 1 $ \cite{Ryang:2006yq}. We also define a new variable,
\begin{eqnarray}
\mathsf{R}=1-r\label{R}
\end{eqnarray}
such that, $ |\mathsf{R}|\ll 1 $ near the UV limit. Finally, re-expressing (\ref{E53}) in terms of this new variable (\ref{R}) we obtain,
\begin{eqnarray}
\Psi \approx T\left(1 + \log (|1-\Lambda|) \right).
\end{eqnarray}

Next, we compute the time difference ($ \Delta \mathfrak{t} $) between the two endpoints of the open string which finally yields\footnote{Here, we rescale the original angle, $ \Delta \mathfrak{t}\rightarrow\Delta \tilde{\mathfrak{t}}=\frac{\Delta \mathfrak{t}}{\upsilon} $.},
\begin{eqnarray}
\cos\frac{\Delta \tilde{\mathfrak{t}}}{2} \approx 1 +\cos \mathcal{D}(\Lambda)\label{E58}
\end{eqnarray}
where,
\begin{eqnarray}
 \mathcal{D}(\Lambda)\approx\log |1-\Lambda |^{1/2}
\end{eqnarray}
is a UV divergent entity and is indeterminant in general as the argument in cosine becomes large in the limit, $ \Lambda \rightarrow 1 $. 

Before we proceed further, it is noteworthy to mention that in the above derivation (\ref{E58}), we have implicitly assumed, $ \upsilon >0 $. If we now further constraints our parameter space such that, $ \omega^{2}_{1}\leq 1-\upsilon^{2} $ and, $ 0<\upsilon <1 $ then combining these constraints with our previous arguments we finally land up into the physical parameter space for our theory,
\begin{eqnarray}
1-\omega_{1}^{2}-\upsilon^{2}\geq 0,~~1-\omega^{2}_{1}\geq \upsilon^{2}\geq 0
\end{eqnarray} 
which thereby precisely matches to that with the parameter bound of the theory in order to be consistent with the three spin giant magnon solution \cite{Ryang:2006yq}.

Next, we compute the remaining conserved quantity,
\begin{eqnarray}
 J_{2}=-2T\int_{\epsilon_{IR}}^{1}\frac{r dr}{(1-r^{2})}.\label{E61}
\end{eqnarray}
Clearly, (\ref{E61}) is IR finite. However, it possesses UV divergences which we need to regularize. To do that, we follow our earlier prescription of UV regularization which finally yields,
\begin{eqnarray}
J_{2}\approx -T(1 -\log|1-\Lambda|).
\end{eqnarray}

Finally, we compute the angle difference between the two end points of the open string,
\begin{eqnarray}
\sin \frac{\Delta \tilde{\varphi}}{2} =-1+\sin \mathcal{D}(\Lambda).
\end{eqnarray} 

Following the original prescriptions \cite{Ryang:2006yq}, we now subtract all the UV divergent pieces and deal only with regularized entities which finally lead towards the following identity,
\begin{eqnarray}
(E-J_{1})_{reg}+\left(\sqrt{\Psi^{2}+\frac{\hat{\lambda}}{4 \pi^{2}}\cos^{2}\frac{\Delta \tilde{\mathfrak{t}}}{2} }\right)_{reg}-\left(\sqrt{J_{2}^{2}+\frac{\hat{\lambda}}{4 \pi^{2}}\sin^{2} \frac{\Delta \tilde{\varphi}}{2}} \right)_{reg}\nonumber\\
 =-\frac{\sqrt{\lambda}\kappa^{2}\omega_{1}}{2 \pi}(1+\kappa^{2})^{1/2}\label{E63}
\end{eqnarray}  
which is the three spin dispersion relation corresponding to $ \kappa $- deformed $ AdS_{3}\times S^{3} $ background in the limit of the large t'Hooft ($ \lambda \gg 1 $) coupling. In the so called undeformed limit ($ \kappa \rightarrow 0 $), Eq.(\ref{E63}) clearly describes the superposition of $ J_{2} $ magnon bound states moving with momentum $ \Delta \tilde{\varphi} $ to that with another bound state of $ \Psi $ magnons moving with momentum, $ \pi +\Delta \tilde{\mathfrak{t}} $  \cite{Ryang:2006yq}. However, in the presence of background deformations, the corresponding interpretation is not so immediate for obvious reasons as explained earlier. Like in the previous example, the above dispersion relation (\ref{E63}) might have some deeper implications from the point of view of the dual gauge theory and therefore needs further attention.

\section{Summary and final remarks}
The present analysis has been devoted towards the understanding of the existing (if any) duality between superstring theories formulated on $ \kappa $- deformed $ AdS_{3}\times S^{3} $ backgrounds to that with their gauge theory counterparts. In order to address this issue, in the first part of our analysis, we perform a detail analysis on the dispersion relation associated with the two spin bound states in the limit of the large t'Hooft coupling ($ \lambda \gg 1 $). Our analysis reveals a non trivial dispersion relation which suggest that the elementary excitation in the dual gauge theory are not the magnons in the usual sense. The \textit{exact} analytic expression corresponding to the two spin magnon excitation could be formally expressed as,
\begin{eqnarray}
\frac{E-J_{1}}{ T}=-\frac{2 \omega  \left(\kappa ^2 r_{0}^2+1\right) (\tan^{2}\mathfrak{j}\kappa )^{1/2}\left(\kappa ^2 \xi ^2 \upsilon ^2+1\right)}{\kappa \Omega (\mathfrak{j}\kappa)e^{-\chi p} }\label{e93}
\end{eqnarray}
where, the exact analytic form corresponding to the functions $ \Omega (\mathfrak{j}\kappa) $ and $ \chi $ have been provided in (\ref{omega}) and (\ref{cai}). In the perturbative ($ \kappa \ll 1 $) regime, the above relation (\ref{e93}) however simplifies to,
\begin{eqnarray}
E-J_{1}-\sqrt{J^{2}_{2}+\frac{\hat{\lambda}}{\pi^{2}}\sin^{2}\frac{\tilde{p}}{2}}=\mathcal{F}(\kappa^{2})=\sqrt{\lambda}\kappa^{2}\left( \frac{\sqrt{2}\mathfrak{j}}{\pi}\right) \mathcal{K}+\mathcal{O}(\kappa^{4})\nonumber\\
\mathcal{K}=2\upsilon^{2}+r^{2}_{0}-\frac{1}{2}-\frac{5}{6}\mathfrak{j}^{2}+\frac{ \mathfrak{b}}{4\mathfrak{j}\sqrt{|1-\mathfrak{j}^{2}|}}.
\end{eqnarray}

In the second part of our analysis, we perform almost identical analysis for spin three bound states and we reach similar conclusions like in the spin two case. The magnon dispersion relation in this case takes the following form,
\begin{eqnarray}
(E-J_{1})_{reg}+\left(\sqrt{\Psi^{2}+\frac{\hat{\lambda}}{4 \pi^{2}}\cos^{2}\frac{\Delta \tilde{\mathfrak{t}}}{2} }\right)_{reg}-\left(\sqrt{J_{2}^{2}+\frac{\hat{\lambda}}{4 \pi^{2}}\sin^{2} \frac{\Delta \tilde{\varphi}}{2}} \right)_{reg}\nonumber\\
 =-\frac{\sqrt{\lambda}\kappa^{2}\omega_{1}}{2 \pi}(1+\kappa^{2})^{1/2}.
\end{eqnarray}  
In summary, the corresponding dual gauge theory interpretation associated with these background deformations is not very clear at this moment and which is therefore worthy of further investigations. 

{\bf {Acknowledgements :}}
This work was supported by the Israeli Science Foundation with Grant Number 1635/16. The author would also like to acknowledge the financial support from the Kreitman School Of Advanced Graduate Studies, The Ben-Gurion University (BGU) of The Negev, Israel.


\begin{thebibliography}{99}

\bibitem{Maldacena:1997re} 
  J.~M.~Maldacena,
  ``The Large N limit of superconformal field theories and supergravity,''
  Int.\ J.\ Theor.\ Phys.\  {\bf 38}, 1113 (1999)
  [Adv.\ Theor.\ Math.\ Phys.\  {\bf 2}, 231 (1998)]
  doi:10.1023/A:1026654312961
  [hep-th/9711200].
  
  \bibitem{Serban:2010sr} 
  D.~Serban,
  ``Integrability and the AdS/CFT correspondence,''
  J.\ Phys.\ A {\bf 44}, 124001 (2011)
  doi:10.1088/1751-8113/44/12/124001
  [arXiv:1003.4214 [hep-th]].
  
  \bibitem{Minahan:2002ve} 
  J.~A.~Minahan and K.~Zarembo,
  ``The Bethe ansatz for N=4 superYang-Mills,''
  JHEP {\bf 0303}, 013 (2003)
  doi:10.1088/1126-6708/2003/03/013
  [hep-th/0212208].
  
  \bibitem{Kruczenski:2003gt} 
  M.~Kruczenski,
  ``Spin chains and string theory,''
  Phys.\ Rev.\ Lett.\  {\bf 93}, 161602 (2004)
  doi:10.1103/PhysRevLett.93.161602
  [hep-th/0311203].
  
  \bibitem{Dimov:2004qv} 
  H.~Dimov and R.~C.~Rashkov,
  ``A Note on spin chain / string duality,''
  Int.\ J.\ Mod.\ Phys.\ A {\bf 20}, 4337 (2005)
  doi:10.1142/S0217751X05020975
  [hep-th/0403121].
  
  \bibitem{Hernandez:2004uw} 
  R.~Hernandez and E.~Lopez,
  ``The SU(3) spin chain sigma model and string theory,''
  JHEP {\bf 0404}, 052 (2004)
  doi:10.1088/1126-6708/2004/04/052
  [hep-th/0403139].
  
   \bibitem{Borsato:2014exa} 
  R.~Borsato, O.~Ohlsson Sax, A.~Sfondrini and B.~Stefanski,
  ``Towards the All-Loop Worldsheet S Matrix for $AdS_3\times S^3\times T^4$,''
  Phys.\ Rev.\ Lett.\  {\bf 113}, no. 13, 131601 (2014)
  doi:10.1103/PhysRevLett.113.131601
  [arXiv:1403.4543 [hep-th]].
  
  \bibitem{Lloyd:2014bsa} 
  T.~Lloyd, O.~Ohlsson Sax, A.~Sfondrini and B.~Stefański, Jr.,
  ``The complete worldsheet S matrix of superstrings on $AdS_3 x S^3 x T^4 $with mixed three-form flux,''
  Nucl.\ Phys.\ B {\bf 891}, 570 (2015)
  doi:10.1016/j.nuclphysb.2014.12.019
  [arXiv:1410.0866 [hep-th]].
  
  \bibitem{Borsato:2016xns} 
  R.~Borsato, O.~Ohlsson Sax, A.~Sfondrini, B.~Stefanski, Jr. and A.~Torrielli,
  ``On the Dressing Factors, Bethe Equations and Yangian Symmetry of Strings on AdS3 x S3 x T4,''
  J.\ Phys.\ A {\bf 50}, no. 2, 024004 (2017)
  doi:10.1088/1751-8121/50/2/024004
  [arXiv:1607.00914 [hep-th]].
  
  \bibitem{Borsato:2013hoa} 
  R.~Borsato, O.~Ohlsson Sax, A.~Sfondrini, B.~Stefanski, Jr. and A.~Torrielli,
  ``Dressing phases of AdS3/CFT2,''
  Phys.\ Rev.\ D {\bf 88}, 066004 (2013)
  doi:10.1103/PhysRevD.88.066004
  [arXiv:1306.2512 [hep-th]].
  
  \bibitem{Borsato:2015mma} 
  R.~Borsato, O.~Ohlsson Sax, A.~Sfondrini and B.~Stefański,
  ``The $\mathrm{AdS}_3\times \mathrm{S}^3\times \mathrm{S}^3\times\mathrm{S}^1$ worldsheet S matrix,''
  J.\ Phys.\ A {\bf 48}, no. 41, 415401 (2015)
  doi:10.1088/1751-8113/48/41/415401
  [arXiv:1506.00218 [hep-th]].
  
  \bibitem{OhlssonSax:2011ms} 
  O.~Ohlsson Sax and B.~Stefanski, Jr.,
  ``Integrability, spin-chains and the AdS3/CFT2 correspondence,''
  JHEP {\bf 1108}, 029 (2011)
  doi:10.1007/JHEP08(2011)029
  [arXiv:1106.2558 [hep-th]].
  
  \bibitem{Babichenko:2009dk} 
  A.~Babichenko, B.~Stefanski, Jr. and K.~Zarembo,
  ``Integrability and the AdS(3)/CFT(2) correspondence,''
  JHEP {\bf 1003}, 058 (2010)
  doi:10.1007/JHEP03(2010)058
  [arXiv:0912.1723 [hep-th]].
  
   \bibitem{Borsato:2016kbm} 
  R.~Borsato, O.~Ohlsson Sax, A.~Sfondrini and B.~Stefański,
  ``On the spectrum of AdS$_3$ × S$^3$ × T$^4$ strings with Ramond–Ramond flux,''
  J.\ Phys.\ A {\bf 49}, no. 41, 41LT03 (2016)
  doi:10.1088/1751-8113/49/41/41LT03
  [arXiv:1605.00518 [hep-th]].
  
  \bibitem{Sax:2012jv} 
  O.~Ohlsson Sax, B.~Stefanski, jr. and A.~Torrielli,
  ``On the massless modes of the AdS3/CFT2 integrable systems,''
  JHEP {\bf 1303}, 109 (2013)
  doi:10.1007/JHEP03(2013)109
  [arXiv:1211.1952 [hep-th]].
    
  \bibitem{Stefanski:2004cw} 
  B.~Stefanski, Jr. and A.~A.~Tseytlin,
  ``Large spin limits of AdS/CFT and generalized Landau-Lifshitz equations,''
  JHEP {\bf 0405}, 042 (2004)
  doi:10.1088/1126-6708/2004/05/042
  [hep-th/0404133].
  
  \bibitem{Hofman:2006xt} 
  D.~M.~Hofman and J.~M.~Maldacena,
  ``Giant Magnons,''
  J.\ Phys.\ A {\bf 39}, 13095 (2006)
  doi:10.1088/0305-4470/39/41/S17
  [hep-th/0604135].
  
  \bibitem{Arutyunov:2006gs} 
  G.~Arutyunov, S.~Frolov and M.~Zamaklar,
  ``Finite-size Effects from Giant Magnons,''
  Nucl.\ Phys.\ B {\bf 778}, 1 (2007)
  doi:10.1016/j.nuclphysb.2006.12.026
  [hep-th/0606126].
  
  \bibitem{Kruczenski:2006pk} 
  M.~Kruczenski, J.~Russo and A.~A.~Tseytlin,
  ``Spiky strings and giant magnons on S**5,''
  JHEP {\bf 0610}, 002 (2006)
  doi:10.1088/1126-6708/2006/10/002
  [hep-th/0607044].
  
 
  \bibitem{Minahan:2006bd} 
  J.~A.~Minahan, A.~Tirziu and A.~A.~Tseytlin,
  ``Infinite spin limit of semiclassical string states,''
  JHEP {\bf 0608}, 049 (2006)
  doi:10.1088/1126-6708/2006/08/049
  [hep-th/0606145].
  
  \bibitem{Spradlin:2006wk} 
  M.~Spradlin and A.~Volovich,
  ``Dressing the Giant Magnon,''
  JHEP {\bf 0610}, 012 (2006)
  doi:10.1088/1126-6708/2006/10/012
  [hep-th/0607009].
  
  \bibitem{Kalousios:2006xy} 
  C.~Kalousios, M.~Spradlin and A.~Volovich,
  ``Dressing the giant magnon II,''
  JHEP {\bf 0703}, 020 (2007)
  doi:10.1088/1126-6708/2007/03/020
  [hep-th/0611033].
  
  \bibitem{Kluson:2007qu} 
  J.~Kluson, R.~R.~Nayak and K.~L.~Panigrahi,
  ``Giant Magnon in NS5-brane Background,''
  JHEP {\bf 0704}, 099 (2007)
  doi:10.1088/1126-6708/2007/04/099
  [hep-th/0703244].
  
  \bibitem{Lukowski:2008eq} 
  T.~Lukowski and O.~Ohlsson Sax,
  ``Finite size giant magnons in the SU(2) x SU(2) sector of AdS(4) x CP**3,''
  JHEP {\bf 0812}, 073 (2008)
  doi:10.1088/1126-6708/2008/12/073
  [arXiv:0810.1246 [hep-th]].
  
  \bibitem{Lee:2008sk} 
  B.~H.~Lee, R.~R.~Nayak, K.~L.~Panigrahi and C.~Park,
  ``On the giant magnon and spike solutions for strings on AdS(3) x S**3,''
  JHEP {\bf 0806}, 065 (2008)
  doi:10.1088/1126-6708/2008/06/065
  [arXiv:0804.2923 [hep-th]].
  
  \bibitem{Lee:2008ui} 
  B.~H.~Lee, K.~L.~Panigrahi and C.~Park,
  ``Spiky Strings on AdS(4) x CP**3,''
  JHEP {\bf 0811}, 066 (2008)
  doi:10.1088/1126-6708/2008/11/066
  [arXiv:0807.2559 [hep-th]].
  
  \bibitem{Lee:2008yq} 
  B.~H.~Lee and C.~Park,
  ``Unbounded Multi Magnon and Spike,''
  J.\ Korean Phys.\ Soc.\  {\bf 57}, 30 (2010)
  doi:10.3938/jkps.57.30
  [arXiv:0812.2727 [hep-th]].
  
  \bibitem{Ahn:2011dq} 
  C.~Ahn and P.~Bozhilov,
  ``Three-point Correlation Function of Giant Magnons in the Lunin-Maldacena background,''
  Phys.\ Rev.\ D {\bf 84}, 126011 (2011)
  doi:10.1103/PhysRevD.84.126011
  [arXiv:1106.5656 [hep-th]].
  
  \bibitem{Bozhilov:2011zp} 
  P.~Bozhilov,
  ``Three-point correlators: Finite-size giant magnons and singlet scalar operators on higher string levels,''
  Nucl.\ Phys.\ B {\bf 855}, 268 (2012)
  doi:10.1016/j.nuclphysb.2011.10.008
  [arXiv:1108.3812 [hep-th]].
  
  \bibitem{Alizadeh:2011yt} 
  D.~Arnaudov and R.~C.~Rashkov,
  ``Three-point correlators: Examples from Lunin-Maldacena background,''
  Phys.\ Rev.\ D {\bf 84}, 086009 (2011)
  doi:10.1103/PhysRevD.84.086009
  [arXiv:1106.4298 [hep-th]].
  
  \bibitem{Ryang:2012pm} 
  S.~Ryang,
  ``Three-point correlator of heavy vertex operators for circular winding strings in $AdS_5 \times S^5$,''
  Phys.\ Lett.\ B {\bf 713}, 122 (2012)
  doi:10.1016/j.physletb.2012.05.049
  [arXiv:1204.3688 [hep-th]].
  
   \bibitem{Ciavarella:2010je} 
  A.~Ciavarella and P.~Bowcock,
  ``Boundary Giant Magnons and Giant Gravitons,''
  JHEP {\bf 1009}, 072 (2010)
  doi:10.1007/JHEP09(2010)072
  [arXiv:1007.1674 [hep-th]].
  
  \bibitem{Bozhilov:2011qf} 
  P.~Bozhilov,
  ``More three-point correlators of giant magnons with finite size,''
  JHEP {\bf 1108}, 121 (2011)
  doi:10.1007/JHEP08(2011)121
  [arXiv:1107.2645 [hep-th]].
  
  \bibitem{David:2008yk} 
  J.~R.~David and B.~Sahoo,
  ``Giant magnons in the D1-D5 system,''
  JHEP {\bf 0807}, 033 (2008)
  doi:10.1088/1126-6708/2008/07/033
  [arXiv:0804.3267 [hep-th]].
  
  \bibitem{Lee:2011fe} 
  B.~H.~Lee and C.~Park,
  ``Finite size effect on the magnon's correlation functions,''
  Phys.\ Rev.\ D {\bf 84}, 086005 (2011)
  doi:10.1103/PhysRevD.84.086005
  [arXiv:1105.3279 [hep-th]].
  
  \bibitem{Park:2010vs} 
  C.~Park and B.~H.~Lee,
  ``Correlation functions of magnon and spike,''
  Phys.\ Rev.\ D {\bf 83}, 126004 (2011)
  doi:10.1103/PhysRevD.83.126004
  [arXiv:1012.3293 [hep-th]].
  
  \bibitem{Hernandez:2011up} 
  R.~Hernandez,
  ``Three-point correlators for giant magnons,''
  JHEP {\bf 1105}, 123 (2011)
  doi:10.1007/JHEP05(2011)123
  [arXiv:1104.1160 [hep-th]].
  
  \bibitem{Arutynov:2014ota} 
  G.~Arutyunov, M.~de Leeuw and S.~J.~van Tongeren,
  ``The exact spectrum and mirror duality of the $(\text{AdS}_5{\times}S^5)_\eta$ superstring,''
  Theor.\ Math.\ Phys.\  {\bf 182}, no. 1, 23 (2015)
  [Teor.\ Mat.\ Fiz.\  {\bf 182}, no. 1, 28 (2014)]
  doi:10.1007/s11232-015-0243-9
  [arXiv:1403.6104 [hep-th]].  
  
  \bibitem{Ali:2015yrs} 
  A.~Mohamed Adam Ali, R.~de Mello Koch, N.~H.~Tahiridimbisoa and A.~Larweh Mahu,
  ``Interacting Double Coset Magnons,''
  Phys.\ Rev.\ D {\bf 93}, no. 6, 065057 (2016)
  doi:10.1103/PhysRevD.93.065057
  [arXiv:1512.05019 [hep-th]].
  
  \bibitem{Ahn:2011zg} 
  C.~Ahn and P.~Bozhilov,
  ``Three-point Correlation functions of Giant magnons with finite size,''
  Phys.\ Lett.\ B {\bf 702}, 286 (2011)
  doi:10.1016/j.physletb.2011.07.011
  [arXiv:1105.3084 [hep-th]].
  
  \bibitem{Beisert:2005tm} 
  N.~Beisert,
  ``The SU(2|2) dynamic S-matrix,''
  Adv.\ Theor.\ Math.\ Phys.\  {\bf 12}, 948 (2008)
  doi:10.4310/ATMP.2008.v12.n5.a1
  [hep-th/0511082].
  
  \bibitem{Dorey:2006dq} 
  N.~Dorey,
  ``Magnon Bound States and the AdS/CFT Correspondence,''
  J.\ Phys.\ A {\bf 39}, 13119 (2006)
  doi:10.1088/0305-4470/39/41/S18
  [hep-th/0604175].
  
  \bibitem{Chen:2006gea} 
  H.~Y.~Chen, N.~Dorey and K.~Okamura,
  ``Dyonic giant magnons,''
  JHEP {\bf 0609}, 024 (2006)
  doi:10.1088/1126-6708/2006/09/024
  [hep-th/0605155].
  
  \bibitem{Bobev:2006fg} 
  N.~P.~Bobev and R.~C.~Rashkov,
  ``Multispin Giant Magnons,''
  Phys.\ Rev.\ D {\bf 74}, 046011 (2006)
  doi:10.1103/PhysRevD.74.046011
  [hep-th/0607018].
  
  \bibitem{Ahn:2010da} 
  C.~Ahn and P.~Bozhilov,
  ``Finite-Size Dyonic Giant Magnons in TsT-transformed $AdS_5\times S^5$,''
  JHEP {\bf 1007}, 048 (2010)
  doi:10.1007/JHEP07(2010)048
  [arXiv:1005.2508 [hep-th]].
  
 \bibitem{Bai:2011su} 
  X.~Bai, B.~H.~Lee and C.~Park,
  ``Correlation function of dyonic strings,''
  Phys.\ Rev.\ D {\bf 84}, 026009 (2011)
  doi:10.1103/PhysRevD.84.026009
  [arXiv:1104.1896 [hep-th]].
  
  \bibitem{Abbott:2009um} 
  M.~C.~Abbott, I.~Aniceto and O.~Ohlsson Sax,
  ``Dyonic Giant Magnons in $CP^3$: Strings and Curves at Finite $J$,''
  Phys.\ Rev.\ D {\bf 80}, 026005 (2009)
  doi:10.1103/PhysRevD.80.026005
  [arXiv:0903.3365 [hep-th]].
  
  \bibitem{Kalousios:2009mp} 
  C.~Kalousios, M.~Spradlin and A.~Volovich,
  ``Dyonic Giant Magnons on CP**3,''
  JHEP {\bf 0907}, 006 (2009)
  doi:10.1088/1126-6708/2009/07/006
  [arXiv:0902.3179 [hep-th]].
  
   \bibitem{Ryang:2006yq} 
  S.~Ryang,
  ``Three-spin giant magnons in AdS(5) x S**5,''
  JHEP {\bf 0612}, 043 (2006)
  doi:10.1088/1126-6708/2006/12/043
  [hep-th/0610037].
  
  \bibitem{Delduc:2013qra} 
  F.~Delduc, M.~Magro and B.~Vicedo,
  ``An integrable deformation of the $AdS_5 \times S^5$ superstring action,''
  Phys.\ Rev.\ Lett.\  {\bf 112}, no. 5, 051601 (2014)
  doi:10.1103/PhysRevLett.112.051601
  [arXiv:1309.5850 [hep-th]].
  
  \bibitem{Hoare:2014pna} 
  B.~Hoare, R.~Roiban and A.~A.~Tseytlin,
  ``On deformations of $AdS_n$ x $S^n$ supercosets,''
  JHEP {\bf 1406}, 002 (2014)
  doi:10.1007/JHEP06(2014)002
  [arXiv:1403.5517 [hep-th]].
  
  \bibitem{Lunin:2014tsa} 
  O.~Lunin, R.~Roiban and A.~A.~Tseytlin,
  ``Supergravity backgrounds for deformations of AdS$_{n} \times S^n$ supercoset string models,''
  Nucl.\ Phys.\ B {\bf 891}, 106 (2015)
  doi:10.1016/j.nuclphysb.2014.12.006
  [arXiv:1411.1066 [hep-th]].
  
  \bibitem{Arutyunov:2014cra} 
  G.~Arutyunov and S.~J.~van Tongeren,
  ``$\mathrm{AdS}_5 \times \mathrm{S}^5$ mirror model as a string sigma model,''
  Phys.\ Rev.\ Lett.\  {\bf 113}, 261605 (2014)
  doi:10.1103/PhysRevLett.113.261605
  [arXiv:1406.2304 [hep-th]].
  
  \bibitem{Arutyunov:2013ega} 
  G.~Arutyunov, R.~Borsato and S.~Frolov,
  ``S-matrix for strings on $\eta$-deformed AdS5 x S5,''
  JHEP {\bf 1404}, 002 (2014)
  doi:10.1007/JHEP04(2014)002
  [arXiv:1312.3542 [hep-th]].
  
  \bibitem{Kameyama:2014vma} 
  T.~Kameyama and K.~Yoshida,
  ``A new coordinate system for $q$-deformed AdS$_{5} \times$ S$^5$ and classical string solutions,''
  J.\ Phys.\ A {\bf 48}, no. 7, 075401 (2015)
  doi:10.1088/1751-8113/48/7/075401
  [arXiv:1408.2189 [hep-th]].
  
  \bibitem{Delduc:2014kha} 
  F.~Delduc, M.~Magro and B.~Vicedo,
  ``Derivation of the action and symmetries of the $q$-deformed $AdS_{5} \times S^{5}$ superstring,''
  JHEP {\bf 1410}, 132 (2014)
  doi:10.1007/JHEP10(2014)132
  [arXiv:1406.6286 [hep-th]].
  
   \bibitem{Khouchen:2014kaa} 
  M.~Khouchen and J.~Kluson,
  ``Giant Magnon on Deformed AdS(3)xS(3),''
  Phys.\ Rev.\ D {\bf 90}, no. 6, 066001 (2014)
  doi:10.1103/PhysRevD.90.066001
  [arXiv:1405.5017 [hep-th]].
  
  \bibitem{Khouchen:2015jfa} 
  M.~Khouchen and J.~Kluson,
  ``D-brane on deformed AdS$_{3} \times$ S$^{3}$,''
  JHEP {\bf 1508}, 046 (2015)
  doi:10.1007/JHEP08(2015)046
  [arXiv:1505.04946 [hep-th]].
  
  \bibitem{Hoare:2014oua} 
  B.~Hoare,
  ``Towards a two-parameter q-deformation of AdS$_3 \times S^3 \times M^4$ superstrings,''
  Nucl.\ Phys.\ B {\bf 891}, 259 (2015)
  doi:10.1016/j.nuclphysb.2014.12.012
  [arXiv:1411.1266 [hep-th]].
  
  \bibitem{Arutyunov:2015mqj} 
  G.~Arutyunov, S.~Frolov, B.~Hoare, R.~Roiban and A.~A.~Tseytlin,
  ``Scale invariance of the $\eta$-deformed $AdS_5\times S^5$ superstring, T-duality and modified type II equations,''
  Nucl.\ Phys.\ B {\bf 903}, 262 (2016)
  doi:10.1016/j.nuclphysb.2015.12.012
  [arXiv:1511.05795 [hep-th]].
  
 
  \bibitem{Banerjee:2015nha} 
  A.~Banerjee, S.~Bhattacharya and K.~L.~Panigrahi,
  ``Spiky strings in $\varkappa$-deformed $AdS$,''
  JHEP {\bf 1506}, 057 (2015)
  doi:10.1007/JHEP06(2015)057
  [arXiv:1503.07447 [hep-th]].
  
  \bibitem{Banerjee:2014bca} 
  A.~Banerjee and K.~L.~Panigrahi,
  ``On the rotating and oscillating strings in (AdS$_{3}$  x S$^{3}$)$_{\kappa}$,''
  JHEP {\bf 1409}, 048 (2014)
  doi:10.1007/JHEP09(2014)048
  [arXiv:1406.3642 [hep-th]].
  
  \bibitem{Banerjee:2016xbb} 
  A.~Banerjee and K.~L.~Panigrahi,
  ``On circular strings in $(AdS_3 \times S^3)_{\varkappa}$,''
  JHEP {\bf 1609}, 061 (2016)
  doi:10.1007/JHEP09(2016)061
  [arXiv:1607.04208 [hep-th]].
  
\bibitem{Kameyama:2015ufa} 
  T.~Kameyama, H.~Kyono, J.~i.~Sakamoto and K.~Yoshida,
  ``Lax pairs on Yang-Baxter deformed backgrounds,''
  JHEP {\bf 1511}, 043 (2015)
  doi:10.1007/JHEP11(2015)043
  [arXiv:1509.00173 [hep-th]].
  
  \bibitem{Kameyama:2016hwl} 
  T.~Kameyama,
  ``Minimal surfaces in q-deformed $AdS_5xS^5$,''
  J.\ Phys.\ Conf.\ Ser.\  {\bf 670}, no. 1, 012028 (2016).
  doi:10.1088/1742-6596/670/1/012028
  
  \bibitem{Klimcik:2015gba} 
  C.~Klimcik,
  ``$\eta$ and $\lambda$ deformations as ${\cal E}$-models,''
  Nucl.\ Phys.\ B {\bf 900}, 259 (2015)
  doi:10.1016/j.nuclphysb.2015.09.011
  [arXiv:1508.05832 [hep-th]].
  
  \bibitem{Arutyunov:2015qva} 
  G.~Arutyunov, R.~Borsato and S.~Frolov,
  ``Puzzles of $\eta$-deformed AdS$_5 \times$ S$^5$,''
  JHEP {\bf 1512}, 049 (2015)
  doi:10.1007/JHEP12(2015)049
  [arXiv:1507.04239 [hep-th]].
  
  \bibitem{Panigrahi:2014sia} 
  K.~L.~Panigrahi, P.~M.~Pradhan and M.~Samal,
  ``Pulsating strings on (AdS$_{3}$ × S$^{3}$)$_{ϰ}$,''
  JHEP {\bf 1503}, 010 (2015)
  doi:10.1007/JHEP03(2015)010
  [arXiv:1412.6936 [hep-th]].
  
  \bibitem{Crichigno:2014ipa} 
  P.~M.~Crichigno, T.~Matsumoto and K.~Yoshida,
  ``Deformations of $T^{1,1}$ as Yang-Baxter sigma models,''
  JHEP {\bf 1412}, 085 (2014)
  doi:10.1007/JHEP12(2014)085
  [arXiv:1406.2249 [hep-th]].
  
  \bibitem{Arutyunov:2014cda} 
  G.~Arutyunov and D.~Medina-Rincon,
  ``Deformed Neumann model from spinning strings on ($AdS_5 \times S^5$)$_\eta$,''
  JHEP {\bf 1410}, 050 (2014)
  doi:10.1007/JHEP10(2014)050
  [arXiv:1406.2536 [hep-th]].
  
  \bibitem{Matsumoto:2014ubv} 
  T.~Matsumoto and K.~Yoshida,
  ``Yang-Baxter deformations and string dualities,''
  JHEP {\bf 1503}, 137 (2015)
  doi:10.1007/JHEP03(2015)137
  [arXiv:1412.3658 [hep-th]].
  
  \bibitem{Lunin:2005jy} 
  O.~Lunin and J.~M.~Maldacena,
  ``Deforming field theories with U(1) x U(1) global symmetry and their gravity duals,''
  JHEP {\bf 0505}, 033 (2005)
  doi:10.1088/1126-6708/2005/05/033
  [hep-th/0502086].
  
  \bibitem{Frolov:2005iq} 
  S.~A.~Frolov, R.~Roiban and A.~A.~Tseytlin,
  ``Gauge-string duality for (non)supersymmetric deformations of N=4 super Yang-Mills theory,''
  Nucl.\ Phys.\ B {\bf 731}, 1 (2005)
  doi:10.1016/j.nuclphysb.2005.10.004
  [hep-th/0507021].
  
  \bibitem{Frolov:2005dj} 
  S.~Frolov,
  ``Lax pair for strings in Lunin-Maldacena background,''
  JHEP {\bf 0505}, 069 (2005)
  doi:10.1088/1126-6708/2005/05/069
  [hep-th/0503201].
  
  \bibitem{Alday:2005ww} 
  L.~F.~Alday, G.~Arutyunov and S.~Frolov,
  ``Green-Schwarz strings in TsT-transformed backgrounds,''
  JHEP {\bf 0606}, 018 (2006)
  doi:10.1088/1126-6708/2006/06/018
  [hep-th/0512253].
  
  \bibitem{Ricci:2007eq} 
  R.~Ricci, A.~A.~Tseytlin and M.~Wolf,
  ``On T-Duality and Integrability for Strings on AdS Backgrounds,''
  JHEP {\bf 0712}, 082 (2007)
  doi:10.1088/1126-6708/2007/12/082
  [arXiv:0711.0707 [hep-th]].
  
  \bibitem{Beisert:2008iq} 
  N.~Beisert, R.~Ricci, A.~A.~Tseytlin and M.~Wolf,
  ``Dual Superconformal Symmetry from AdS(5) x S**5 Superstring Integrability,''
  Phys.\ Rev.\ D {\bf 78}, 126004 (2008)
  doi:10.1103/PhysRevD.78.126004
  [arXiv:0807.3228 [hep-th]].
  
 \end{thebibliography}
\end{document}